\documentstyle[12pt]{article}
\newcommand{\beq}{\begin{equation}}
\newcommand{\ber}{\begin{eqnarray}}
\newcommand{\eeq}{\end{equation}}
\newcommand{\eer}{\end{eqnarray}}
\begin{document}
\rightline{[SUNY BING 10/30/96]}
\vspace{1mm}
\begin{center}
{\LARGE \bf Tau Decays Beyond the Standard Model}\\
\end{center}
\vspace{10mm}
\noindent
{Charles A. Nelson\footnote{Electronic address:
cnelson@bingvmb.cc.binghamton.edu \newline Invited talk at
TAU96 workshop}, Department of Physics, State
University of New York at Binghamton, Binghamton, N.Y.
13902-6016}\\
\vspace{10mm}
\begin{abstract}
In a recent paper, 8 semileptonic parameters were defined
to specify the most general Lorentz-invariant spin
correlation functions for 2-body $\tau$ decays. These
parameters can be used to search for anomalous $\Gamma_L$,
$\Gamma_T$ polarized-partial widths, for non-CKM-type
leptonic
CP violation, and for leptonic $ \tilde{T}_{FS} $ violation.
They can also be used to bound the effective-mass
scales $ \Lambda $ for ``new physics" arising from additional
Lorentz structures, e.g. from lepton compositeness, tau weak
magnetism, weak electricity, or second-class currents. It is
emphasized that (i) for these tests ``different modes have
different merits"
and that (ii) the parameters can be measured either by using
spin-correlation techniques without polarized beams, or with
longitudinally polarized beams.
\end{abstract}
\vskip 40 pt

\section{INTRODUCTION}

Currently, the bounds are very weak for possible ``new
physics" in $\tau$ decays.  At best, the limits are at the
several percent level whereas the errors on the Michel
paratmeters are typically at the
per-mill level in $\mu$ decays .  During the last five years,
impressive high precision electoweak experiments have been
performed at the $Z$ boson resonance at LEP and at the
SLC. From these experiments, and those by ARGUS, BES, and
CLEO, information on $\tau$ decays at the several-percent
level has been obtained.
Now, the time has come for high precision experiments in
$\tau$ decays. Model independent analyses, which {\bf do not
assume} a mixture of $(V \mp A)$ couplings, are now necessary
in
$\tau$ decays as a means for searching for ``new physics"
beyond the standard model.

For the purely leptonic decays, $\tau^- \rightarrow l^-
\bar{\nu_l} \nu_{\tau}$ and $\tau^+ \rightarrow l^+ \nu_l
\bar{ \nu_{\tau} }$, there is the classic Michel
parametrization.  This has been reviewed in Pich's talk
\cite{pich}.  Here we will concentrate on \newline 2-body
$\tau$ decays
for which the branching ratios are large.  In a recent paper
\cite{can1}, 8 semileptonic parameters were defined to
specify the most general Lorentz-invariant spin correlation
functions for 2-body $\tau$ decays.  They can also be
measured
at an $e^- e^+$ collider with longitudinally-polarized beams,
such as at the SLC or at a future tau/charm factory\cite{a1}.

Conclusions of this talk include:
\begin{itemize}
    \item ``Different modes have different merits" in
systematic searchs for new physics.
    \item Even without candidates for lepton-number violating
decays or for other forbidden
modes, new physics can be discovered in
on-going and future experiments by exploiting $\tau^- \tau^+$
spin-correlations and/or longitudinal-beam polarization.
\end{itemize}

Recent papers by theorists on possible ``new
physics" in $\tau$ decays, include studies of $\Delta L_i
\neq
0$ at future colliders \cite{hall}, of leptonic
$CP$-violation
\cite{kilian,kil2,kil4} and of special effects in the
third family
\cite{reina}. Leptoquark and SUSY  mechanisms have been
proposed for producing observable $CP$-violating dipole
moments in $\tau^- \tau^+$ production \cite{m9}. The
forthcoming BNL experiment has motivated a more precise
treatment of higher-order hadronic contributions to the
anomalous magnetic moments of the $\mu$ and $\tau$
in the standard model\cite{1}.

\section{TESTS FOR ``NEW PHYSICS"}

By using a general formalism for two-body $\tau$
decays, one can (a) determine the ``complete Lorentz
structure" of ${J^{Charged}}_{Lepton}$ directly from
experiment, and (b) test in a model independent manner for
the
presence of ``new physics".  For instance, there are
simple tests for non-CKM-type leptonic $CP$ violation and for
leptonic $ \tilde{T}_{FS} $ violation in $\tau$ decays.

\subsection{General formalism for two-body \newline $\tau$
decays:}

The physical idea is very simple:  We introduce 8 parameters
to describe the most
general spin-correlation function for the decay sequence $
Z^o,\gamma
^{*}\rightarrow \tau ^{-}\tau ^{+}\rightarrow (\rho ^{-}\nu
)(\rho ^{+}\bar
\nu )$\ followed by $\rho ^{ch}\rightarrow \pi
^{ch}\pi ^o$ including both
$\nu _L$,\ $\nu _R\ $helicities and both   $\bar \nu _R$,\
$\bar \nu _L\
$helicities.  We present the
discussion for the $\rho
\nu$ channel, but the
same formulas hold for the $a_1 \nu$ and $K^* \nu$ channels.
Thus, by including the $\rho $ polarimetry
information that is available from the $%
\rho^{ch}\rightarrow \pi^{ch}\pi ^o$ decay distribution, the
polarized-partial-widths for $\tau ^{-}\rightarrow
\rho ^{-}\nu \ $ are directly measureable.  For instance,
the general angular distribution for
polarized $\tau^{-}_{L,R} \rightarrow \rho ^{-}\nu
\rightarrow
(\pi ^{-}\pi
^o)\nu $ is described by
\begin{eqnarray}
  {dN} / d(\cos \theta_1^\tau )
d(\cos \tilde \theta_a) d \tilde \phi_a
= {\bf n}_a [1\pm {\bf f}_a \cos
\theta _1^\tau ] \nonumber \\
\nonumber \\
\mp
(1/\sqrt{%
2})\sin \theta _1^\tau \sin 2\tilde \theta _a\ {\cal R}_\rho
[\omega \cos
\tilde \phi
_a+\eta ^{\prime }\sin \tilde \phi _a]
\end{eqnarray}
with upper(lower) signs for a L-handed $\tau^-$ \newline
(R-handed),
where
\begin{equation}
\begin{array}{c}
{\bf n}_a=\frac 1{8}(3+\cos 2\tilde \theta _a+\sigma {\cal
S}_\rho [1+3\cos
2\tilde
\theta _a]) \\ {\bf n}_a{\bf f}_a=\frac 1{8}(\xi [1+3\cos
2\tilde
\theta _a]+\zeta {\cal S}_\rho [3+\cos 2\tilde \theta _a])
\end{array}
\end{equation}
In this expression, $\cos \theta_1^\tau $ describes the
direction of the $\rho^-$ momentum in the $\tau^-$ rest
frame, and $\cos \tilde \theta _a$, and $\tilde \phi_a$
describe the direction of the $\pi^-$ in the $\rho^-$ rest
frame. Such formulas for more general spin-correlation
functions in terms of the 8 semi-leptonic parameters are
given in [2] for unpolarized $e^- e^+$ beams, and in [14] for
polarized beams. \newline

There are eight $\tau$ semi-leptonic decay parameters since
there are the four $\rho_{L,T} \nu_{L,R}$ final states:
The {\bf first parameter} is simply $\Gamma \equiv
\Gamma
_L^{+}+\Gamma _T^{+}$, i.e. the (full) partial width
for $\tau ^{-}\rightarrow \rho ^{-}\nu $.  The subscripts on
the $\Gamma $'s denote the
polarization of the final $\rho ^{-}$ (and in the SM of the
intermediate off-shell
$W^-$ boson), either
``L=longitudinal'' or
``T=transverse''; superscripts denote ``$\pm $ for
sum/difference of
the $\nu _{L\ }$versus $\nu _R$ contributions''
\begin{equation}
\begin{array}{c}
\Gamma _L^{\pm }=\left| A(0,-\frac 12)\right| ^2\pm \left|
A(0,\frac
12)\right| ^2 \\
\Gamma _T^{\pm }=\left| A(-1,-\frac 12)\right| ^2\pm \left|
A(1,\frac
12)\right| ^2.
\end{array}
\end{equation}

The {\bf second} is the
chirality
parameter $\xi \equiv \frac 1\Gamma (\Gamma _L^{-}+\Gamma
_T^{-})$.
Equivalently, \newline \newline $ \xi \equiv$ (Prob
$\nu_{\tau}$ is
$\nu_L$) $ - $
(Prob $\nu_{\tau}$ is $\nu_R$),
\begin{equation}
\hspace{1pc}  \equiv |< \nu_L |\nu_{\tau} >|^{2} - |< \nu_R
|\nu_{\tau}
>|^{2}
\end{equation}
So a value $\xi = 1$ means the coupled $\nu_{\tau}$ is
pure $\nu_L$.   The {\bf
remaining two partial-width parameters} are defined by
\begin{equation}
\zeta \equiv (\Gamma _L^{-}-\Gamma _T^{-})/(
{\cal S}_\rho \Gamma ), \hspace{1pc} \sigma \equiv (\Gamma
_L^{+}-\Gamma
_T^{+})/(
{\cal S}_\rho \Gamma ).
\end{equation}
The definiton for $\sigma$ implies that
\newline  \hspace{2pc} $ \tilde{\sigma} \equiv {\cal S}_\rho
\sigma =$ (Prob
$\rho $ is $\rho _L$)
$
-
$ (Prob $\rho $ is $\rho _T$), \newline is the analogue of
the
neutrino's chirality parameter in
Eq.(4).
Thus, the parameter $\tilde{\sigma}$ measures
the
degree of polarization of the emitted $\rho$.  If the
exchange
is completely via an off-shell $W$-boson, $\tilde{\sigma}$
measures
the polarization of the $W$-boson.

The interference between these ${\rho / W }_L$ and
${\rho / W }_R$ amplitudes can be determined by measuring the
{\bf four parameters},
\begin{equation}
\begin{array}{c}
\omega \equiv I_{
{\cal R}}^{-}\ /({\cal R}_\rho \Gamma ), \hspace{2pc}  \eta
\equiv I_{
{\cal R}}^{+}\ /({\cal R}_\rho \Gamma ) \\ \omega ^{\prime
}\equiv I_{
{\cal I}}^{-}\ /({\cal R}_\rho \Gamma ), \hspace{2pc} \eta
^{\prime }\equiv
I_{{\cal I}%
}^{+}\ /({\cal R}_\rho \Gamma )
\end{array}
\end{equation}
The associated $LT$-interference intensities are
\begin{equation}
\begin{array}{c}
I_{{\cal R}}^{\pm }=\left| A(0,-\frac 12)\right| \left| A(-
1,-\frac
12)\right| \cos \beta _a  \cr \pm \left| A(0,\frac
12)\right|
\left| A(1,\frac
12)\right| \cos \beta _a^R  \\
I_{{\cal I}}^{\pm }=\left| A(0,-\frac 12)\right| \left| A(-
1,-\frac
12)\right| \sin \beta _a \cr \pm \left| A(0,\frac
12)\right|
\left| A(1,\frac
12)\right| \sin \beta _a^R
\end{array}
\end{equation}
Here $\beta _a\equiv \phi _{-1}^a-\phi _0^a$, and $\beta
_a^R\equiv \phi
_1^a-\phi _0^{aR}$\ are the measurable phase differences of
of the
associated helicity amplitudes
$A(\lambda_{\rho},\lambda_{\nu})=\left|
A\right| \exp \iota \phi $. \newline

In the standard lepton model in which there is only a $(V-A)$
coupling and $m_{\nu} = 0$, these parameters all equal
one
except that the two parameters directly sensitive to leptonic
$ \tilde{T}_{FS} $ violation vanish, ${\omega ^{\prime } } =
{\eta ^{\prime }
} = 0$.  Note that {\bf in the \newline special case} of a
mixture of
only $V$ \& $A$
couplings and $m_{
\nu_{\tau} } = 0 $, $\xi \rightarrow \frac{\left| g_L\right|
^2-\left|
g_R\right| ^2}{\left| g_L\right| ^2+\left|g_R\right| ^2}$ so
that the ``stage-one spin correlation" parameter $\zeta
\rightarrow \xi$.  Thus, in this special case $\zeta$
measures
the $\nu_{\tau}$ helicity and  $\zeta = \xi$, {\bf but for
more general
couplings
neither property holds} for $\zeta$ for
$\tau \rightarrow \rho \nu$, $a_1 \nu$, $K^* \nu$.

Hadronic factors ${\cal S}_\rho $ and ${\cal R}_\rho
$,
\begin{equation}
{\cal S}_\rho =\frac{1-2\frac{m_\rho
^2}{m^2}}{1+2\frac{m_\rho ^2}{m^2}} ,
\hspace{2pc}
{\cal R}_\rho =\frac{\sqrt{2}\frac{m_\rho }m}{1+2\frac{m_\rho
^2}{m^2}} .
\end{equation}
have been explicitly inserted into the definitions of some of
the semi-leptonic decay parameters, so that quantities such
as
${q_\rho }^2={m_\rho }^2$ can be
smeared over in application due to the finite $\rho $ width.
These factors numerically are $({\cal
S},{\cal R})_{\rho
,a_1,K^{*}}=0.454,0.445;-0.015,0.500;0.330,0.472$.

From Table 1 given below for the $\Gamma_L$, $\Gamma_T$
polarized-partial-widths, one easily sees
that the numerical values of ``$\xi, \zeta, \sigma, \ldots $"
are very different for unique Lorentz couplings. This is
indicative of the analyzing power of polarization techniques
in two-body $\tau$ decay modes. Both the real and
the imaginary parts of the associated helicity amplitudes can
be directly measured, c.f. Eqs(3,7).

\subsection{Tests for non-CKM-CP and \newline $
\tilde{T}_{FS} $
violations:}

These formulas only assume Lorentz invariance and do not
assume any
discrete symmetry properties.  Therefore, it is easy to use
this framework for
testing for discrete symmetry properties.  In particular,
with $A\left( \lambda _\rho ,\lambda _\nu
\right)$ for  $%
\tau ^{-}\rightarrow \rho ^{-}\nu $  and with $B\left(
\lambda _{\bar \rho },\lambda _{\bar \nu }\right)$ for $\tau
^{+}\rightarrow
\rho
^{+}\bar
\nu $ a specific discrete symmetry implies a specific
relation among the associated
helicity amplitudes:%
$$
\begin{array}{cc}
\underline{Invariance} & \underline{Relation} \\ P & A\left(
-
\lambda _\rho
,-\lambda _\nu \right) =A\left( \lambda _\rho ,\lambda _\nu
\right)  \\
& B\left( -\lambda _{\bar \rho },-\lambda _{\bar \nu }\right)
=B\left(
\lambda _{\bar \rho },\lambda _{\bar \nu }\right)  \\
C & B\left( \lambda _{\bar \rho },\lambda _{\bar \nu }\right)
=A\left(
\lambda _{\bar \rho },\lambda _{\bar \nu }\right)  \\
CP & B\left( \lambda _{\bar \rho },\lambda _{\bar \nu
}\right)
=A\left(
-\lambda _{\bar \rho },-\lambda _{\bar \nu }\right)  \\
\tilde T_{FS} & A^{*}\left( \lambda _\rho ,\lambda _\nu
\right)
=A\left(
\lambda _\rho ,\lambda _\nu \right)  \\
& B^{*}\left( \lambda _{\bar \rho },\lambda _{\bar \nu
}\right)
=B\left(
\lambda _{\bar \rho },\lambda _{\bar \nu }\right)  \\
CP\tilde T_{FS} & B^{*}\left( \lambda _{\bar \rho },\lambda
_{\bar
\nu
}\right) =A\left( -\lambda _{\bar \rho },-\lambda _{\bar \nu
}\right)
\end{array}
$$
In the  $\tau ^{-}$ rest frame, the
matrix element for $\tau ^{-}\rightarrow \rho ^{-}\nu$ is
$
\langle \theta _1^\tau ,\phi _1^\tau ,\lambda _\rho ,\lambda
_\nu
|\frac
12,\lambda _1\rangle =D_{\lambda _1,\mu }^{\frac 12*}(\phi
_1^\tau ,\theta
_1^\tau ,0)A\left( \lambda _\rho ,\lambda _\nu \right)
$
where $\mu =\lambda _\rho -\lambda _\nu $ and $\lambda_1$ is
the $\tau^{-}$
helicity.   For the $CP$-conjugate
process, $\tau
^{+}\rightarrow \rho ^{+}\bar \nu \rightarrow \left( \pi
^{+}\pi
^o\right) \bar \nu $, in the $\tau ^{+}$ rest frame
$
\langle \theta _2^\tau ,\phi _2^\tau ,\lambda _{\bar \rho
},\lambda
_{\bar
\nu }|\frac 12,\lambda _2\rangle =D_{\lambda _2,\bar \mu
}^{\frac
12*}(\phi
_2^\tau ,\theta _2^\tau ,0)B\left( \lambda _{\bar \rho
},\lambda
_{\bar \nu
}\right)
$
with $\bar \mu =\lambda _{\bar \rho }-\lambda _{\bar \nu }$.
\newline

{\bf ``Measurement of a non-real helicity
amplitude implies
a violation of $\tilde T_{FS}$ invariance"}. This true since
by exact $T$ invariance the
amplitudes for the
process of interest and the time reversed process must be
equal:
\begin{equation}
{\cal M}(\tau ^{-}\rightarrow \rho ^{-} \nu_\tau )={\cal
N(}\tau
^{-}\leftarrow \rho ^{-}\nu _\tau )
\end{equation}
This can be written as
\begin{equation}
\langle \lambda _\rho \lambda _\nu |H_{eff}|\lambda _\tau
\rangle =\langle
\lambda _\tau |H_{eff}|\lambda _\rho \lambda _\nu \rangle
\end{equation}
in terms of the ``effective Hamiltonian'' $H_{eff}$ which
describes the
transition. In quantum mechanics, the right-hand-side equals
the complex
conjugate matrix element
\begin{equation}
\langle \lambda _\tau |H_{eff}|\lambda _\rho \lambda _\nu
\rangle =\overline{%
\langle \lambda _\rho \lambda _\nu |H_{eff}^{\dagger
}|\lambda
_\tau \rangle
}
\end{equation}
but with the initial and final states interchanged and
Hermitian-adjoint $%
H_{eff}^{\dagger}$. {\bf If} {\bf there are no} \newline
``final state
interactions'', the
effective Hamiltonian will be Hermitian
$H_{eff}=H_{eff}^{\dagger }$ so
\begin{equation}
\langle \lambda _\rho \lambda _\nu |H_{eff}|\lambda _\tau
\rangle =\overline{%
\langle \lambda _\rho \lambda _\nu |H_{eff}|\lambda _\tau
\rangle }
\end{equation}
or
\begin{equation}
{\cal M}(\tau ^{-}\rightarrow \rho ^{-}\nu _\tau )={\cal
M}(\tau
^{-}\rightarrow \rho ^{-}\nu _\tau )^{*}
\end{equation}
Therefore, for an exclusive tau decay mode, the associated
transition
amplitude will be real by canonical $T$-invariance if
there are no \newline ``final state interactions.'' {\bf We
refer to this
as $\tilde
T_{FS}$-invariance.}

A violation of $\tilde T_{FS}$-invariance could
occur because of the exchange of a Z$%
^{^{\prime }}$ boson between the final $\rho ^{-
}$and
the final $%
\nu _\tau $ in which the Z$^{^{\prime }}$ couples differently
to
the $\rho _L$
versus the $\rho _T$. Or $\tilde T_{FS}$-violation
could occur
because of a fundamental violation of canonical
$T$-invariance. Whatever the
cause might turn out to be, the experimental discovery of a
violation of $%
\tilde T_{FS}$-invariance in a tau two body decay mode would
be very significant. \newline

In this formalism,
\begin{itemize}
    \item If the primed parameters $ \omega ^{\prime } \neq 0
$ and/or $
\eta
^{\prime }
\neq 0
\Longrightarrow  $ {\bf $\tilde{T}_{FS} $ is violated:}
\end{itemize}

In this formalism there are four parameters $\eta ,\eta
^{^{\prime }},\omega ,\omega ^{^{\prime
}}$ which can be used to test for leptonic $ \tilde{T}_{FS} $
violation.  This occurs because the trigonometric structure
of Eqs.(7)
implies the two constraints
\begin{equation}
(\tilde{\eta} \pm \tilde{\omega} )^2+(\tilde{\eta^{\prime }}
\pm
\tilde{\omega^{\prime }} )^2=\frac
14[(1 \pm \xi )^2-( \tilde{\sigma} \pm \tilde{\zeta} )^2].
\end{equation}
among these four parameters.

To test for leptonic
$ \tilde{T}_{FS} $ violation, besides the $\omega$
parameter which can be measured from $I_4$ in both the $\rho$
and $a_1$
modes, there is the $\eta ^{\prime } $ parameter which can be
obtained
from $I_5$ in both the $\rho$ and $a_1$ modes.  Also there
are
the $\eta$ and $\omega ^{\prime } $ parameters which only
appear in
S2SC
distributions for the $a_1$ modes.  The $a_1$ mode parameters
are more sensitive [2]
for searching for leptonic $ \tilde{T}_{FS} $ violation than
the simple $I_4$
distribution considered in Ref.[11].

{\bf Canonical CPT invariance} implies only equal total
widths between a
particle and its antiparticle.  Canonical CPT invariance does
not imply equal partial widths between CP-conjugate decay
modes of a particle and its antiparticle.  Indeed, in nature
in the
kaon system the {\bf partial widths} of the neutral kaons do
differ
for
the particle and the antiparticle. \newline

The barred parameters $ \bar{\xi},
\bar{\zeta}, \ldots $ have
the analogous
definitions [2] for the CP conjugate modes, $\tau
^{+}\rightarrow
\rho
^{+}\bar{\nu}, \ldots $. For instance,

$ \bar \xi =$ (Prob $\bar \nu _\tau $
is $\bar \nu
_R$)
$ - $ (Prob $\bar \nu _\tau $ is $\bar \nu _L$) ,
\begin{equation}
 \hspace{2pc} \bar \Gamma _L^{\pm }=|B(0,\frac 12|^2\pm
|B(0,-\frac
12|^2.
\end{equation}
Therefore, in this formalism
\begin{itemize}
    \item If any $
\bar{\xi} \neq \xi,
\bar{\zeta} \neq \zeta, \ldots $ $ \Longrightarrow $ {\bf CP
is
violated:}
\end{itemize}

As was shown in [5], if
only $\nu_L$ and $\bar\nu_R$ exist,  there are two simple
tests for ``non-CKM-type" leptonic CP violation in $\tau
\rightarrow \rho \nu$ decay.  Normally a CKM leptonic-phase
will contribute equally at tree level to both the $\tau^-$
decay amplitudes and so will cancel out in the ratio of their
moduli and in their relative phase (for exceptions see
footnotes
14, 15 in [5]).  The two tests for leptonic CP
violation are: $
\beta _a=\beta _b $, where $\beta _a=\phi _{-1}^a-\phi _0^a$,
$\beta _b=\phi _1^b-
\phi
_0^b$, and
$
r_a=r_b $, where $
r_a=\frac{|A\left( -1,-\frac 12\right) |}{|A\left( 0,-\frac
12\right)
|}$, $ r_b=\frac{|B\left( 1,\frac 12\right) |}{|B\left(
0,\frac
12\right) |} $.
Sensitivity levels for $\tau \rightarrow \rho \nu$ and
$\tau \rightarrow  a_1 \nu$ decays are to about $0.05$ to $
0.1\%$ for $r_a = r_b$, and to about $1^o$ to $3^o$ for
$\beta_a = \beta_b$ at $10 GeV$ and at $4 GeV$ without using
polarized $e^- e^+$ beams [5, 2].

\begin{table*}[htb]
\setlength{\tabcolsep}{2.pc}
\newlength{\digitwidth} \settowidth{\digitwidth}{\rm 0}
\catcode`?=\active \def?{\kern\digitwidth}
\caption{Tests for anomalous $\Gamma_L$, $\Gamma_T$
polarized-partial-widths for $\rho_{L,T} \nu_{L,R}$ final
states:}
\label{tabpre3}
\begin{tabular*}{\textwidth}{@{}lllll}
\hline
\cline{2-3} \cline{4-5}
                 & \multicolumn{1}{l}{$V \mp A$}
                 & \multicolumn{1}{l}{$S \pm P$}
                 & \multicolumn{1}{l}{$f_M + f_E$}
                 & \multicolumn{1}{l}{$f_M - f_E$}         \\
\hline
{$\bf  \; \; Analytic$}                    &
&

&

&

\\
$\Gamma _L^{-} / \Gamma$
& $\pm \frac{1}{2} (1 + {\cal S}_\rho ) $
&
$\pm 1
$            &  $\frac{ \rho^2 }{2 \tau^2 + \rho^2}$
&
$ - \frac{1}{3}
$
\\
$\Gamma _T^{-} / \Gamma$                           & $\pm
\frac{1}{2} (1 - {\cal S}_\rho )  $
&
$ 0
$           & $ \frac{ 2 \tau^2 }{2 \tau^2 + \rho^2}$
&
 $ - \frac{2}{3}                                 $
\\
$\Gamma _L^{+} / \Gamma$                          & $
\frac{1}{2} (1 + {\cal S}_\rho )$
&
$ 1
$            &  $\frac{ \rho^2 }{2 \tau^2 + \rho^2}$
&
$    + \frac{1}{3}                                 $
\\
$\Gamma _T^{+} / \Gamma$                          & $
\frac{1}{2} (1 - {\cal S}_\rho )$
&
$ 0
$            &  $\frac{ 2 \tau^2 }{2 \tau^2 + \rho^2}$
&
$    + \frac{2}{3}                                 $
\\
{$\bf  \; \; Numerical$}       &
&

&

&

\\
$\Gamma _L^{-} / \Gamma$                              & $\pm
0.7(\pm 0.5)$ & $
\pm 1  $

& $                  0.0(0.2)
$     &
 $ -0.3                             $
\\
$\Gamma _T^{-} / \Gamma$                              & $\pm
0.3(\pm 0.5)$      &  $
0   $

&  $1.0(0.8)
$
&
$-0.7$
\\
$\Gamma _L^{+} / \Gamma$                               & $
0.7(0.5)$
&  $ 1
$
& $ 0.0(0.2)
$         &
$ +0.3 $
\\
$\Gamma _T^{+} / \Gamma$                                & $
0.3(0.5)$
&  $ 0
$
&  $ 1.0(0.8)
$         &
$ +0.7 $
\\
\hline
\multicolumn{5}{@{}p{160mm}}{Values for unique Lorentz
couplings. ${\cal S}_\rho $ is defined in Eq.($8$).  Entries
are for $\rho^-$ ( $a_1^-$, if value
differs).}
\end{tabular*}
\end{table*}

\subsection{Tests for anomalous $\Gamma_L$, $\Gamma_T$
polarized-partial-widths:}

The contribution of the longitudinal($L$) and transverse($T$)
$\rho$/$W$ amplitudes in the decay
process is projected out by the simple formulas:
\newline
\begin{eqnarray*}
I_{
{\cal R}}^{\nu _L,\nu _R}\equiv \frac 12(I_{{\cal R}}^{+}\pm
I_{{\cal R}%
}^{-})=|A(0,\mp \frac 12||A(\mp 1,\mp \frac 12|\cos \beta
_a^{L,R} \\ =\frac
\Gamma 2(\tilde{\eta} \pm \tilde{\omega} )
\end{eqnarray*}
\begin{eqnarray*}
I_{
{\cal I}}^{\nu _L,\nu _R}\equiv \frac 12(I_{{\cal I}}^{+}\pm
I_{{\cal I}%
}^{-})=|A(0,\mp \frac 12||A(\mp 1,\mp \frac 12|\sin \beta
_a^{L,R} \\ =\frac
\Gamma 2(\tilde{\eta^{\prime }} \pm \tilde{\omega^{\prime }}
)
\end{eqnarray*}
\begin{eqnarray*}
\Gamma _L^{\nu _L,\nu _R}\equiv \frac 12(I_L^{+}\pm I_L^{-
})=|A(0,\mp \frac
12|^2 \\
=\frac \Gamma 4(1+\tilde{\sigma} \pm \xi \pm \tilde{\zeta} )
\end{eqnarray*}
\begin{eqnarray*}
\Gamma _T^{\nu _L,\nu _R}\equiv \frac 12(I_T^{+}\pm I_T^{-
})=|A(\mp 1,\mp
\frac 12|^2 \\
=\frac \Gamma 4(1-\tilde{\sigma} \pm \xi \mp \tilde{\zeta} )
\end{eqnarray*}
In the first line, $\beta _a^L=\beta _a$ but we normally
suppress such L
superscripts, e.g. Eq.(7). Unitarity, requires the two right-
triangle
relations
$$
\begin{array}{c}
(I_{
{\cal R}}^{\nu _L})^2+(I_{{\cal I}}^{\nu _L})^2=\Gamma
_L^{\nu
_L}\Gamma
_T^{\nu _L} \\

(I_{{\cal R}}^{\nu _R})^2+(I_{{\cal I}}^{\nu
_R})^2=\Gamma
_L^{\nu _R}\Gamma _T^{\nu _R}.
\end{array}
$$

Table 1 gives the analytic forms and numerical values of the
$\Gamma_L$, $\Gamma_T$ partial-widths for unique Lorentz
couplings.  An important experimental goal is to determine
whether or not these partial widths are anomalous in nature
versus the standard lepton model's $(V-A)$ predictions
(single Higgs doublet) because the
$W_L$ versus $W_T$ partial widths might have distinct
dynamical differences if electroweak dynamical symmetry
breaking occurs in nature.

\subsection{Tests for additional Lorentz \newline
structures:}

Besides model independence, a major open issue is
whether
or not there is an additional chiral coupling in the tau's
charged-current. A chiral classification of
additional structure is a natural phenomenological extension
of the
symmetries of the standard $SU(2)_L\ X\ U(1)$\ electroweak
lepton model.
For \hskip1em  $\tau^{-
}\rightarrow \rho
^{-}\nu _{L,R}$, the most general Lorentz coupling is
\begin{equation}
\rho _\mu ^{*}\bar u_{\nu _\tau }\left( p\right) \Gamma ^\mu
u_\tau
\left(
k\right)
\end{equation}
where $k_\tau =q_\rho +p_\nu $. It is convenient to treat the
vector
and
axial vector matrix elements separately. In Eq.(16)
\begin{eqnarray*}
\Gamma _V^\mu =g_V\gamma ^\mu +
\frac{f_M}{2\Lambda }\iota \sigma ^{\mu \nu }(k-p)_\nu   +
\frac{g_{S^{-}}}{2\Lambda }(k-p)^\mu  \\
+\frac{g_S}{2\Lambda
}(k+p)^\mu
+%
\frac{g_{T^{+}}}{2\Lambda }\iota \sigma ^{\mu \nu }(k+p)_\nu
\end{eqnarray*}
\begin{eqnarray*}
\Gamma _A^\mu =g_A\gamma ^\mu \gamma _5+
\frac{f_E}{2\Lambda }\iota \sigma ^{\mu \nu }(k-p)_\nu \gamma
_5
+
\frac{g_{P^{-}}}{2\Lambda }(k-p)^\mu \gamma
_5  \\
+\frac{g_P}{2\Lambda }%
(k+p)^\mu \gamma _5  +\frac{g_{T_5^{+}}}{2\Lambda }\iota
\sigma ^{\mu \nu
}(k+p)_\nu \gamma _5
\end{eqnarray*}

The parameter
$%
\Lambda =$ ``the effective-mass scale of new physics''. In
effective field
theory
this
is the scale at which new particle thresholds are expected to
occur or where the theory becomes non-perturbatively
strongly-interacting so as to overcome perturbative
inconsistencies.  It can also be interpreted as a measure of
a
new compositeness scale.  In old-fashioned renormalization
theory
$\Lambda$  is the scale at
which the calculational methods and/or the principles of
``renormalization''
breakdown.

Without additional theoretical or
experimental
inputs, it is not possible to select what is the ``best"
minimal set of couplings for
analyzing the structure of the tau's charged current.  For
instance, by  Lorentz
invariance, for the $\rho$, $a_1$, $K^*$ modes there are the
equivalence theorems that for the
vector (axial-vector)
current
\begin{eqnarray}
S\approx V+f_M, & T^{+}\approx -V+S^{-}
\end{eqnarray}
\begin{eqnarray}
P\approx -A+f_E, & T_5^{+}\approx A+P^{-}
\end{eqnarray}
There are similar but different equivalences for the $\pi$,
$K$ modes, see Eq.(21). Therefore, from the perspective of
searching for the fundamental dynamics, it
is important to investigate what limits can be set for a
variety of Lorentz structures (including $S^{\pm}$,
$P^{\pm}$, $T^{\pm}$, and ${T_5}^{\pm}$) and not just for a
kinematically minimal, but theoretically prejudiced, set.

Ref.\cite{2} gives the limits on $\Lambda$ in $GeV$ for real
$g_i$'s from the $\rho$ and $a_1$ modes:  Effective
mass scales of $\Lambda \sim 1-2 TeV$ can be probed at $10
GeV$, and at $4 GeV$ for the $(S+P)$ and the $f_M + f_E$
couplings. For determination of ideal statistical errors, we
assume  $10^7$ ($\tau ^{-}\tau ^{+}$%
) pairs at $10GeV$ and separately at $4GeV$; at $M_Z$ we
assume $10^7$
$Z^o$'s with BR($Z^o\rightarrow \tau ^{-}\tau
^{+}$)$=0.03355$; $BR_{\rho}=24.6 \%$, $BR_{a_1}=18\%$ for
the sum of neutral/charged $a_1$ modes, and $BR_{\pi}=11.9
\%$.

We listed the ideal
statistical error for the presence of an additional $V+A$
coupling as an error $\delta (\xi _A)$ on the chirality
parameter $\xi _A$
for $\tau ^{-}\rightarrow A^{-}\nu $. Equivalently, if one
ignores possible
different L and R leptonic CKM factors, the effective lower
bound on an
additional $W_R^{\pm }$ boson (which couples only to right-
handed
currents)
is
$$
M_R=\{\delta (\xi _A)/2\}^{-1/4}M_L
$$
For the $\{\rho ^{-},\rho ^{+}\}$($\{a_1{}^{-},a_1{}^{+}\}$)
mode, from $\delta (\xi _\rho
)=0.0012(0.0018)$ this gives equivalently
$M_R>514GeV(464GeV)$. Probably, $10^8$($\tau ^{-
}\tau ^{+}$)
pairs will be accumulated by a $\tau $/charm factory at
$4GeV$, so all
the potential $4GeV$ statistical-error bounds might be
improved by a factor of
$3.2$.

\section{DIFFERENT MODES HAVE \hspace*{2cm}
DIFFERENT MERITS}

In contrast to the purely leptonic modes\cite{pich,6}, the
tau
semi-leptonic modes are qualitatively distinct since they
enable a second-stage spin-correlation.  From existing
results, a quantitative comparison
with the ideal sensitivity
in the purely leptonic case is possible if we assume an
arbitrary mixture of
$V$ and $A$ couplings with $m_\nu =0.$ Then the semi-leptonic
chirality parameter $%
\xi _\rho $ and the chiral polarization parameter $\xi
_{Lepton}$ can be
compared since then they both equal $(|g_L|^2-
|g_R|^2)/(|g_L|^2+|g_R|^2)$.
By using $I_4$ to obtain $\xi _\rho $ from $\{\rho ^{-}\rho
^{+}\}$, the
statistical error [2] is $\delta (\xi _\rho )=0.006$ at
$M_Z$.
This is a
factor of $8$ better than the pure leptonic mode's $\delta
(\xi
_{Lepton})=0.05$ error\cite{5} from averaging over the $\mu $
and
$e$ modes and
using $I_3(E_1,E_2,\cos \psi _{12})$ where $\psi_{12}$ is the
openning angle between the two final charged leptons in the
cm-
frame. A complete determination
of the purely
leptonic parameters for $\tau ^{-}\rightarrow \mu ^{-}
\bar{\nu
_\mu} \nu _\tau $
will require a difficult measurement of the $\mu $
polarization,
see
Fetscher[12].

\subsection{$\rho$, $K^*$ modes:}

For the $\rho^-$ mode, the errors for ($\xi, \zeta, \sigma,
\omega$)
based on simple four-variable spin-correlation function
$I_4$ are slightly less than $ 1 \% $: For $10^7$ ($\tau^- ,
\tau^+ $) pairs at 10 GeV:
from the $ \lbrace \rho^- , \rho^+ \rbrace $ mode and using
the
four-variable distribution $I_4$, the ideal statistical
percentage errors are for $\xi$, $0.6 \% $; for $ \zeta$, $
0.7
\% $; for $ \sigma$, $ 1.3 \% $; and for $ \omega$, $ 0.6 \%
$.  The
CP tests for these semileptonic parameters are about $\sqrt2$
worse.  Typically the $a_1$ values for these parameters are
about 3 times worse
than the $\rho$ values.

In analogy with the Pauli anomalous magnetic moment, an
obvious signature for lepton compositeness would be an
additional tensorial coupling. In this regard, it is useful
to first test for the presence of only $\nu_L$ couplings
which would exclude a significant contribution from the $g_{-
} = f_M - f_E$ tensorial coupling:
\begin{itemize}
    \item For the $a_1$
and $\rho$ modes there are 3
logically independent tests for only $\nu_L$ couplings:
\newline $\xi
= 1$, $\zeta = \sigma$, and $\omega = \eta$.
\end{itemize}
In addition, if
$\tilde T_{FS}$ violation
occurred then the non-zero parameters $\omega^{'} = \eta^{'}$
if there are only $\nu_L$ couplings. An additional tensorial
$g_{+}=f_M+f_E$ coupling
would preserve these 3 signatures for only $\nu _L$
couplings.  But, such a tensorial $g_{+}$ coupling
would give them non-($V-A$)-values: $\zeta = \sigma \neq 1$
and
$\omega = \eta \neq 1$. Second, for a  $g_{+}$ coupling there
is the prediction that for $\Lambda $ large
\begin{equation}
(\zeta -1)=(1-\omega )\frac gl
\end{equation}
where  the ratio `` $g/l$ '' is a known function [2] of
$m_\rho $ and $%
m_\tau $. Numerically  $(g/l)_{\rho }=0.079$.

These $\nu_L$ signatures and Eq.(19) also
occur for an additional $(S+P)$ coupling but with
the ratio $(g/l)$ replaced by $(a/d)$, which varies from
$5.07$ to $12.1$ across
$(m_{\rho} \pm \Gamma / 2 $.  Fortunately, here the $\pi$
mode can again be used to limit the presence of an additional
$(S+P)$ coupling versus a $g_+ = f_M + f_E$ coupling since
the latter does not contribute to the $\pi$ mode.

\subsection{$a_1$ mode:}

For the kinematic description of $\tau ^{-}\rightarrow a_1^{-
}\nu
\rightarrow (\pi _1^{-}\pi _2^{-}\pi _3^{+})\nu $\ , the
normal to the $(\pi
_1^{-}\pi _2^{-}\pi _3^{+})\ $decay triangle is used in place
of the $\pi
^{-}$\ momentum direction of the $\tau ^{-}\rightarrow \rho
^{-}\nu
\rightarrow (\pi ^{-}\pi ^o)\nu \ $ sequential decay.

Including both $\nu _L\ $and $\nu _R\ $ helicities, the
composite
decay density matrix for $\tau ^{-}\rightarrow a_1^{-}\nu
\rightarrow (\pi
_1^{-}\pi _2^{-}\pi _3^{+})\nu \ $ is%
\begin{equation}
{\bf R}^\nu =S_1^{+}{\bf R}^{+}+S_1^{-}{\bf R}^{-}
\end{equation}
where the sequential decay density matrices ${\bf R}^{\pm }$\
describing $\tau ^{-}\rightarrow a_1^{-
}\nu
\rightarrow (\pi _1^{-}\pi _2^{-}\pi _3^{+})\nu $\ are given
in Ref.[2].
The  $S_1^{\pm }\ $ factors do depend on the
strong-interaction form-factors used to
describe the decay $a_1^{-}\rightarrow \pi _1^{-}\pi _2^{-
}\pi _3^{+}$. However, when the 3-body Dalitz plot is
integrated over, only
the $S_1^{+}\ $%
term remains, so it can be absorbed into the overall
normalization factor
which removes any arbitrary form-factor dependence.
Similarly, by asymmetric integration, the $S_1^{-}\ $%
factor can be absorbed into an overall
normalization factor.

To test for leptonic $\tilde T_{FS}$-violation, besides the
$\omega$
parameter which can be measured from $I_4$ in both the $\rho$
and $a_1$
modes, there are the following from the $\{a_1^-,a_1^+\}$
mode:
using $I_5^{-}$ the errors are
for
$ \eta $, $0.6 \% $; using $I_7 $ for $ \eta^{'} $, $ 0.013
$; and using $ I_7^{-} $ for $ \omega{'} $, $ 0.002 $.  See
Sec.(2.2) above.

\subsection{$\pi$, $K$ modes:}

The $\tau ^{-}\rightarrow
\pi ^{-}\nu ,K^{-}\nu $ modes
each generally provide less
information since here only two of the semi-leptonic
parameters can be measured,i.e. the partial widths
$\Gamma_{\pi,K}$ and the chirality parameter $\xi _{\pi ,K}
=\frac{\left| A(-\frac
12)\right| ^2-\left|
A(\frac 12)\right| ^2}{\left| A(-\frac 12)\right| ^2+\left|
A(\frac
12)\right| ^2}$.
Second, the important
weak-magnetism and weak-electricity couplings, $f_M$ and
$f_E$, do not contribute to these modes; but, they can be
precisely measured by the $\rho$ and $a_1$ modes.  Third, by
Lorentz invariance, there are the strong equivalence theorems
\begin{equation}
S^{-}\approx S\approx T^{+}\approx V , P^{-}\approx
P\approx
T_5^{+}\approx A
\end{equation}
because there are only two independent decay amplitudes.

Nevertheless, from the $\pi $ mode there is
good separation ($>127 GeV$ from CLEO II data) of $V-A$ from
a $T^{+}+T_5^{+}$ coupling, whereas these couplings cannot be
separated in the
$\rho$ and $a_1$ modes. Second, the $S+P$ coupling is also
excluded to $\Lambda >127 GeV$.  Third, there is direct
measurement
of
the chirality parameter $\xi_{\pi}$, i.e. of
the probablity that the emitted $\nu_{\tau}$ is L-handed.
Unfortunately, the fundamental $S^{-}$\ and
$P^{-}$\ couplings which do not contribute to $\tau
\rightarrow
\rho \nu ,a_1\nu ,K^{*}\nu $ are suppressed in
$\tau ^{-}\rightarrow
\pi ^{-}\nu
,K^{-}\nu $\ decay since
$q\cdot V\sim \frac{%
m_\pi ^2}{2\Lambda }g_{S^{-}}$\ and $q\cdot A\sim \frac{m_\pi
^2}{2\Lambda }%
g_{P^{-}}$.

It is also important to note what cannot be precisely
measured by two-body $\tau$ decay modes:  \newline (i) The
present
and potential experimental bounds on $(S^- \pm P^-)$
couplings are exceptionally poor or non-existent from
measurements of the $\pi$, $\rho$ and $a_1$ modes[11, 2].
(ii)
The S2SC functions do not enable a measurement of
any relative phase between the $\nu_L$ and $\nu_R$ helicity
amplitudes[2].

\section{TESTS WITH LONGITUDINALLY-POLARIZED BEAMS}

For the case of longitudinally
polarized beams, we assume 100 \% polarization and study
the 3-variable distribution ${I_3}^{\cal P} ( \theta_{beam},
E_{\rho^-}, \tilde{\theta}_{\pi^{-}} )$.  In the center-of-
mass frame,  $\theta_{beam}$ is the angle between the final
tau momentum and the initial $e^-$ beam, and  $E_{\rho^-}$ is
the energy of the final $\rho^-$.  The angle
$\tilde{\theta}_{\pi^{-}}$ is the direction of the final
$\pi^-$ momentum in the $\rho^-$ rest frame [when boost is
directly from the center-of-mass frame]. We call this the ``
$\cal{P}_L$ method".  Instead of the angle between the $e^-$
and $\tau^-$ momenta, one could use the angle between the
$e^-$ and the final $\rho^-$ momenta; work on this
alternative 3-variable distribution in in progress\cite{4} .

See Table 2 for the errors for measurement of the ($\xi,
\zeta, \sigma, \omega$) parameters
based on $I_3^{\cal{P}}$ and on $I_4$. In general, by using
longitudinally-polarized beams the errors for the $\rho^-$
mode are slightly less than $ 0.4 \% $ and about a factor of
7 better than by using the S2SC function $I_4$.  The
CP tests for these semileptonic parameters are $\sqrt2$
worse by the $\cal{P}_L$ method, or by the
S2SC method.  Typically the $a_1$ values are 2-4 times worse
than the $\rho$ values. However, for $\xi$, the error for the
$a_1$ mode by the $\cal{P}_L$ method is about 3 times better
than that for the $\rho$ mode.

Both methods are
comparable for the two tests for non-CKM-type leptonic $CP$
violation.  Table 3 shows the sensitivities of the
$\rho$ and $a_1$ modes.

For helpful discussions, we thank participants at this
conference.  This work was partially supported by U.S. Dept.
of Energy Contract No. DE-FG 02-96ER40291.

{\bf For copy of last two tables send email to author.}

\end{document}